# EQUATIONS OF MOTION OF THE MASS CENTERS IN A SCALAR THEORY OF GRAVITATION: EXPANSION IN THE SEPARATION PARAMETER [*]

MAYEUL ARMINJON

*Laboratoire "Sols, Solides, Structures"*
*[CNRS / Université Joseph Fourier / Institut National Polytechnique de Grenoble]*
*B.P. 53, 38041 Grenoble cedex 9, France.*
*E-mail: arminjon@hmg.inpg.fr*

*Abstract.* An asymptotic framework is defined for the small parameter $\eta$ that quantifies a good separation between the extended bodies that make a weakly gravitating system. This is introduced within an alternative scalar theory of gravitation, though it may be defined similarly in other theories. This framework allows one to truncate the translational equations of motion at any well-defined order. Here, the post-Newtonian (PN) equations valid in the scalar theory are truncated beyond the order $\eta^3$. The PN approximation scheme used is the asymptotic scheme, that expands all fields. To get the explicit form of the equations of motion for the mass centers, the bodies are assumed spherical, merely for calculating the PN corrections. It is found that, due to the use of the asymptotic PN scheme, the internal structure of the bodies does play a role in the equations of motion.

Short Title: *Motion of well-separated bodies in a scalar theory of gravitation*

## 1. INTRODUCTION

According to its current practice, relativistic celestial mechanics in the solar system consists in summing a Newtonian contribution and a relativistic contribution, the latter depending on the assumption that the moving bodies behave like point masses. Indeed, the ephemerides for the main bodies of the solar system are currently based either [1-2] on the Lorentz-Droste-Einstein-Infeld-Hoffmann (LDEIH) equations [3-5], or instead [6-7] on the equations of motion of a test particle in a Schwarzschild field, and, in both cases, one has to add Newtonian contributions that account for figure effects [1-2, 6-7] (and also for the perturbations by the planets, for the latter ephemerides [6-7]). The derivations of small general-relativistic effects, such as the Fokker and Schiff precessions or the Lense-Thirring effect, consider point masses in the field of one massive body, whose spatial extension and material structure do not appear in the derivations (see *e.g.*

---

Refs. 8-11). It is clearly desirable to account for the finite extension of the bodies that constitute a self-gravitating system, because one should investigate in detail what are the effects of the structure of the bodies, and of their internal motions, on the orbits that they follow according to a relativistic theory of gravitation. (Such effects are to be expected as a consequence of the mass-energy equivalence.) Following Fock [12] and Papapetrou [13], a number of authors have developed post-Newtonian (PN) approximation schemes for extended bodies in general relativity (GR). But, in most of these works, the really explicit equations of motion with which one ends up, if any, are just the LDEIH equations – the latter being recovered because some strong simplifications are further made, not because they would apply to a general system of extended bodies at the first PN (1PN) approximation. (The latter is indeed not true, *e.g.* because the spin does play a role, see the quotation of Refs. [15-18] below.) Even in the thorough work of Fock [14], the final form of the 1PN equations of motion for a system of extended bodies is given a bit indirectly, by writing a Lagrangian which is the sum of seven terms, each of which fills a sizeable equation. Among the few examples (known to the author) of explicit translational equations of motion, derived for a system of extended bodies at the 1PN approximation of GR, and which do generalize the LDEIH equations, there are those obtained by Damour *et al.* [15], that include "spin-orbit" and "spin-spin" terms in addition to the LDEIH contribution, and that extend equations previously obtained by Tulczyjew [16] and by Barker and O'Connell [17]. And those of Spyrou [18], in which the difference with the LDEIH equations is also related to the spin.

Furthermore, besides the DSX formalism [5,15], there has rather recently appeared in GR yet another PN approximation scheme [19] (see also Ref. 20). From the viewpoint of modern approximation theory, this scheme [19] looks much more natural than the standard one [9,14,21], in that the scheme [19] is a rather straightforward application of the usual method of asymptotic expansion for a system of partial differential equations. Essentially, the new scheme [19] introduces a one-parameter family of initial-value problems, thus defining a family of solution fields (in Ref. 20, a family of solution fields is *a priori* assumed, without reference to an initial-value problem). However, the initial conditions for the spatial metric are very particular in Ref. 19; it could be difficult to extend the approach [19] to a general-enough system, because in GR the initial conditions for the spatial metric have to verify the nonlinear *constraint equations*. Whereas local equations in this "asymptotic" scheme have been written for some gauge conditions [19-20] and some mathematical properties of the solutions have been studied [19-20], no equivalent even of the LDEIH equations has yet been given for the asymptotic scheme of GR. On the other hand, equations of motion of the mass centers, including spin effects, have been obtained [22-23] within the asymptotic scheme proposed [24] for an alternative, scalar theory of gravitation. The scalar character of that theory simplifies the calculations significantly as compared with GR. Hence, in addition to the interest that one may find in testing that scalar ether-theory (see Ref. 25 for a motivation and a summary of the latter, which will be named hereafter "the scalar theory"), a detailed investigation of the PN equations of motion according to the asymptotic scheme in this theory may also be considered as a preparation for a similar study in GR. Let us recall that, also because the scalar theory is simpler than GR, it is easier, in the scalar theory, to *compare* the standard PN scheme with the "asymptotic" one, and this comparison shows that the standard scheme indeed is not consistent with the usual method of asymptotic expansion [24]. Hence it would seem very important to develop the use of the asymptotic scheme *in GR also,* up to the obtainment of the equations of motion of the mass centers. However, it could be difficult to do this for a general-enough system, due to the above-mentioned difficulty with the initial conditions.

In the previous work, the most general form of the translational 1PN equations of motion (of the scalar theory) has been given [22], and it has been made tractable in two successive simplification steps [23]. Yet it has been found since that the second simplification step was too drastic, in that it dropped out some terms which must be retained in the solar system. This lack of the work [23] (despite the fact that no error has been seen) came, as we shall see here, from the *absence of an asymptotic framework for the separation parameter $\eta$* introduced there [$\eta$ is defined by Eq. (2.1) below]. *The aim of the present paper* is: **i**) to define a such asymptotic framework for $\eta$ ; **ii**) to use this framework so as to exploit the good separation between bodies in a more satisfying way than in (all) previous attempts, by consistently truncating the expansions with respect to $\eta$ ; **iii**) to obtain the explicit equations for a truncation order which will be shown appropriate in the solar system. In *Section 2*, we shall introduce a definite asymptotic framework for the separation parameter $\eta$. *Section 3* will point out one compelling reason why some terms, which were neglected for working out the final form of the equations in Ref. 23, must actually be retained for calculations in the solar system. This will enforce us to keep all terms up to and including $\eta^3$ in the translational equations of motion. In *Section 4*, we shall give the final form of the equations for the 1PN corrections to the motion of the mass centers, as expanded up to and including the order $\eta^3$. (The necessary calculations are summarized in an *Appendix.*) As in the former work [23], the 1PN *corrections,* but not necessarily the zero-order contributions themselves, will assume that the zero-order mass density, hence also the zero-order pressure and the Newtonian self-potential, are spherically symmetric for each body.[1] *Section 5* will show the derivation of initial conditions for the translational 1PN corrections. Our main conclusions will be given in *Section 6*. We shall refer to the previous work [22-23], and shall use the corresponding notations. Henceforth, "PN" will usually mean "1PN".

## 2. ASYMPTOTIC FRAMEWORK FOR WELL-SEPARATED BODIES

---

[1] This distinction is important, because the zero-order (Newtonian) effects of the departure from sphericity, while very small as compared with the Newtonian main terms for the main bodies of the solar system, have a magnitude comparable to that of the 1PN corrections. Thus, in the solar system, it seems fully justified to neglect the departure from sphericity when calculating the 1PN corrections, but it would be an error to do so for the zero-order contribution. Now, in all PN schemes but the asymptotic scheme, the 1PN positions and velocities are given in one piece (see *e.g.* the LDEIH equations), not as the sum of a zero-order term and a 1PN correction. This means that, except in the asymptotic scheme, one cannot consider the departure from sphericity in a self-consistent way, unless one does so in the whole system of PN equations – which would be very complicated.

In the solar system, the distances between bodies are much larger than their sizes. This leads to define a *separation parameter* thus:

(2.1) $\qquad \eta_0 \equiv \text{Max}_{a \ne b} (r_b / |\mathbf{a} - \mathbf{b}|), \qquad r_b \equiv \tfrac{1}{2} \text{Sup}_{\mathbf{x},\mathbf{y} \in D_b} |\mathbf{x} - \mathbf{y}|.$

(We use the zero-order positions $\mathbf{a}$, $\mathbf{b}$, … of the mass centers of the $N$ bodies $(a)$, $(b)$, …, with $a$, $b$, … $= 1, …, N$: since the PN corrections to these positions are very small, this is immaterial for the definition of the small parameter $\eta_0$. $D_a$ is the time-dependent domain occupied by the generic body $(a)$ in the *preferred reference frame* of the theory.) Now we wish to use Taylor expansions, *e.g.* at the order 1:

$$(2.2) \qquad \frac{1}{R} \equiv \frac{1}{|\mathbf{X} - \mathbf{x}|} = \frac{1}{|\mathbf{X} - \mathbf{b}|} + \frac{(\mathbf{x} - \mathbf{b}).(\mathbf{X} - \mathbf{b})}{|\mathbf{X} - \mathbf{b}|^3} + S(\mathbf{x}, \mathbf{X}),$$

$$(2.3) \qquad S(\mathbf{x}, \mathbf{X}) \le \frac{1}{2} |x^i - b^i| |x^j - b^j| \text{Sup}_{0 \le \zeta \le 1} \left| \frac{\partial^2 (1/|\mathbf{Y}|)}{\partial Y^i \partial Y^j} \right|_{\mathbf{Y} = \mathbf{X} - \mathbf{b} + \zeta(\mathbf{b} - \mathbf{x})}.$$

To deduce consistent estimates, *we must introduce an asymptotic framework for the small parameter $\eta$. To do the latter thing, we shall now define a family $(S^\eta)$ of well-separated PN systems, all analog with our physically given system S, and with the gravitational field having the same order of magnitude as in this system, but in which the parameter $\eta$ has an arbitrary value.* The idea is to expand PN fields and equations as $\eta \to 0$, and then to use these equations for the finite small value $\eta_0$ valid for the physically given system S. To the author's knowledge, no such family of well-separated systems has ever been considered in previous attempts at rigorously accounting, within GR or any other relativistic theory of gravity, for the "good separation" between bodies. (This good separation does occur, *e.g.*, in the solar system.) For example, no such family is considered in Ref. 18. In Ref. 26 (not known to the author for the previous version of this work), a family of well-separated systems is considered given in the study of the *Newtonian* problem of motion (pp. 137-145 in Ref. 26), although the family is not explicitly defined. But no family of well-separated systems is considered in Ref. 26 at the stage of studying the *post-Newtonian* approximation (of GR). As it will appear in Sect. 3 and in the Appendix, the lack, in our previous work [23], of considering the small parameter $\eta$ within a such definite asymptotic framework (based on a conceptual family of well-separated systems) led then to the inappropriate neglect of some numerically significant terms in the equations of motion at the first PN approximation of the scalar theory.

Essentially, we shall consider that the system $S^\eta$ is made of bodies which are identical to those of the system S, but that their separation distances are of order $\eta^{-1}$ exactly:

(2.4) $\qquad | \mathbf{a}^\eta - \mathbf{b}^\eta | = \text{ord}(\eta^{-1}).$

Since the fields are defined as solutions of an initial-value problem [24], we actually have to define the independent fields at the initial time only. Because our starting equations here are the first-order equations of the asymptotic PN approximation, which form a closed system of equations (see §§ 2.3 and 2.4 in Ref. 22), it is enough to define the PN fields. (The asymptotic PN equations are not physically self-sufficient, however, in that they do not ensure by themselves that the weak-field/ low-velocity assumptions are satisfied.) The PN gravitational fields, *i.e.* the Newtonian potential associated with the zero-order rest-mass density $\rho$: $U \equiv$ N.P.[$\rho$], plus the PN field $A$, which is determined by the fields $B$ and $W$ [22, Eqs. (4.14-15)], depend merely on the PN matter fields. Since the initial values of the latter ones depend only [22, Eqs. (2.32-34)] on the initial *zero-order* matter fields: the 0-order pressure $p(T = 0)$ [or equivalently the density $\rho(T = 0)$], and the 0-order velocity $\mathbf{u}(T = 0)$, we just have to define the latter fields. We set

(2.5) $$\mathbf{a}^\eta(T = 0) = \mathbf{a}(T = 0)\, \eta_0/\eta,$$

(2.6) $$\rho^\eta(\mathbf{x}, T = 0) = \rho(\mathbf{a} + \mathbf{y}, T = 0) \quad \text{if } \mathbf{x} = \mathbf{a}^\eta + \mathbf{y} \text{ with } \mathbf{a} + \mathbf{y} \in D_a(T = 0).$$

Equation (2.6) defines $\rho^\eta$ so that, indeed, the density inside the bodies is independent of $\eta$ [setting $\rho^\eta(\mathbf{x}, T = 0) = 0$ if $\mathbf{x}$ does not have the form above for some $a = 1, ..., N$], whereas (2.5) ensures that (2.4) is satisfied, at least near $T = 0$. To define the velocity $\mathbf{u}^\eta(T = 0)$, we use the auxiliary assumption that each body undergoes a rigid motion at the Newtonian approximation [14]:

(2.7) $$u^i = \dot{a}^i + \Omega^{(a)}{}_{ji}\,(x^j - a^j), \text{ or } \mathbf{u} = \dot{\mathbf{a}} + \boldsymbol{\omega}_a \wedge (\mathbf{x} - \mathbf{a}), \text{ for } \mathbf{x} \in D_a, \quad (\Omega^{(a)}{}_{ji} + \Omega^{(a)}{}_{ij} = 0).$$

Of course, this is only approximately true (*e.g.* due to the tidal influence of the other bodies), but we shall use this assumption merely to calculate the PN corrections. Since the latter ones are very small, it certainly implies only an extremely small error in the solar system. Anyhow, we can assume that (2.7) is exact at the initial time. From the Newtonian estimate $\dot{\mathbf{a}}^2 \approx U^{(a)}(\mathbf{a})$, valid in the reference frame of the global mass center, and from (2.4), we expect that, at any time,

(2.8) $$(\dot{a}^i)^\eta = \mathrm{ord}(\eta^{1/2}).$$

We shall assume that this is true if we define the initial translation velocities of system $S^\eta$ as

(2.9) $$(\dot{a}^i)^\eta(T = 0) = (\eta/\eta_0)^{1/2}\, \dot{a}^i(T = 0).$$

As to the self-rotation velocities, in the solar system they have at most the same magnitude, in linear values, as the translation velocities, and our numerical calculations show that the PN corrections containing quadratic terms in $\Omega^{(a)}{}_{ji}$, included in the final translational equations of Ref. 23, are negligibly small in the solar system. To avoid such terms in the expansions, it turns out to be sufficient that

(2.10) $$(\Omega^{(a)}{}_{ji})^\eta \ll \eta^{1/2}.$$

Hence we shall set, for some small number $\varepsilon > 0$,

(2.11) $$(\Omega^{(a)}{}_{ji})^\eta (T=0) = (\eta/\eta_0)^{\varepsilon + 1/2}\, \Omega^{(a)}{}_{ji}(T=0),$$

and we shall assume that this ensures that (2.10) is true at any time.

### 3. THE APPROPRIATE TRUNCATION ORDER IN $\eta$ IN THE SOLAR SYSTEM

The final translational equations of Ref. 23 {the Newtonian equation, plus the equation for the 1PN corrections, Eq. (2.20) in Ref. 23a or Eq. (6.17) in Ref. 23b, hereafter referred to as "the equations [23]"}, were numerically implemented. The motion of Mercury around the Sun, as obtained by numerical integration of the equations of motion of a test particle in a Schwarzschild field [27], was compared with the motion of Mercury around the Sun, as deduced from the numerical solution of the equations [23] for the two-bodies system Sun-Mercury. The "asymptotic" version of the equations of motion of a test particle in a Schwarzschild field [27]:

(3.1) $$\mathbf{x}_{\text{exact}} = \mathbf{x} + \mathbf{x}_1/c^2 + O(1/c^4), \qquad \mathbf{u}_{\text{exact}} = \mathbf{u} + \mathbf{u}_1/c^2 + O(1/c^4),$$

(3.2) $$\frac{d\mathbf{u}}{dt} = -\frac{GM}{r^2}\mathbf{n} \qquad (\mathbf{n} \equiv \mathbf{x}/|\mathbf{x}|,\quad r \equiv |\mathbf{x}|),$$

(3.3) $$\frac{d\mathbf{u}_1}{dt} = \frac{GM}{r^2} \times \left\{ \left( \frac{2GM}{r} - \frac{M_1}{M} + 3\left[ (\mathbf{u}\cdot\mathbf{n})^2 + \frac{\mathbf{x}_1\cdot\mathbf{n}}{r} \right] - 2\mathbf{u}^2 \right)\mathbf{n} + 2(\mathbf{u}\cdot\mathbf{n})\mathbf{u} - \frac{\mathbf{x}_1}{r} \right\},$$

was used for this comparison, since the equations [23] are based on the asymptotic method of PN expansion. [2] [In (3.2) and (3.3), $M$ and $M_1$ are the coefficients of the first-order asymptotic expansion of the active mass (of the Sun, here) in Schwarzschild's exterior solution.] It was found that these two motions differ in a surprisingly significant manner from one another, if one remembers that the investigated scalar theory predicts Schwarzschild's exterior solution in the spherical static case. Unsurprisingly, this great difference could be traced back to the differences in the accelerations. In particular, in the "asymptotic Schwarzschild" equations of motion, the PN correction (3.3) to the acceleration contains terms that are linear in the PN correction $\mathbf{x}_1$ to the position:

---

[2] However, the asymptotic and standard methods of PN approximation can be made numerically equivalent in the case of a test particle in a Schwarzschild field [27].

$$
(3.4) \qquad \left(\frac{d\mathbf{u}_1}{dt}\right)_{\mathbf{x}_1} \equiv \frac{GM}{r^2} \times \left(3\frac{\mathbf{x}_1 \cdot \mathbf{n}}{r}\mathbf{n} - \frac{\mathbf{x}_1}{r}\right),
$$

and it was found that these terms play a numerically significant role – whereas the equations [23] do not contain any such term. Since the equations [23] are a simplification of Eq. (4.1) below, this means that the simplification must have unappropriately eliminated such terms. And since $I^{ai}$ and $K^{ai}$ in Eq. (4.1) depend merely on the zero-order fields [22, Eqs. (4.8) and (4.10)], such terms, *i.e.* terms that are linear in the vector $\mathbf{a}_1$, which allows to calculate the PN correction to the position of the mass center, have to come from the integral $J^{ai}$. The first simplification of this integral [23a, Eq. (2.17)] kept two groups of terms, the ratio of these two groups being $O(\eta)$, and only the second (higher-order) group contains $\mathbf{a}_1$. Thus we may guess that it is this second group which gives the equivalent of (3.4), and indeed this second group was eliminated in the second simplification step [23a, Eq. (3.12)]. It has been verified that keeping this second group gives just the term (3.4) in the point particle limit, and it has been found that it is of order $\eta^3$ (see after Eq. (A14) in the Appendix). Therefore, we must keep all terms up to and including (at least) the order $\eta^3$ for calculations in the solar system. Fortunately, the order $\eta^3$ turns out to be enough (see the end of Sect. 5).

## 4. THE PN EQUATIONS OF MOTION OF THE MASS CENTERS AT THE ORDER $\eta^3$ (SPHERICAL BODIES)

The general form of the equation for the PN corrections to the motion of the mass centers is [22, Eq. (4.9)]:

$$
(4.1) \qquad M_a^1 \ddot{a}_1^i + \dot{I}^{ai} = J^{ai} + K^{ai},
$$

where $M_a^1$, $I^{ai}$, $J^{ai}$ and $K^{ai}$ are defined as integrals of PN fields. Our main task is to get these integrals in a tractable form, using as in Ref. 23 the assumption that the zero-order density $\rho$ is spherically symmetrical for each body (see Note 1 here), *and* using now the asymptotic framework introduced in Sect. 2. This is a straightforward modification of the previous calculations [23b]. This modification is summarized in the Appendix here. Then, the explicit form of the PN correction to the translational equations of motion is got by entering the obtained expressions of the four integrals [Eqs. (A10, A14, A18-20, A24-28)] into (4.1). In space vector form, this may be written as

$$
(4.2) \quad M_a^1 \ddot{\mathbf{a}}_1 = o(\eta^3) + \left[\left(\tau_a - \tfrac{5}{6} M_a\right)\dot{\mathbf{a}}^2 - \tfrac{5}{3} M_a U^{(a)}(\mathbf{a}) - \tfrac{17}{3}\varepsilon_a - 2T_a\right]\nabla U^{(a)}(\mathbf{a})
$$

$$
-\left[\left(\tau_a + 2M_a\right)\dot{\mathbf{a}}\cdot\nabla U^{(a)}(\mathbf{a}) + \tfrac{4}{3} GM_a \sum_{b\neq a} M_b \frac{\mathbf{n}_{ab}\cdot\dot{\mathbf{b}}}{r_{ab}^2}\right]\dot{\mathbf{a}} + \left[2\varepsilon_a + M_a U^{(a)}(\mathbf{a})\right]\boldsymbol{\omega}_a \wedge \dot{\mathbf{a}}
$$

$$+ G \sum_{b \neq a} \left\{ \frac{M_a M_b}{2 r_{ab}} \left[ (\mathbf{n}_{ab} \cdot \ddot{\mathbf{b}}) \mathbf{n}_{ab} - \ddot{\mathbf{b}} \right] + \frac{1}{r_{ab}^2} \left[ -\alpha_{ab} \mathbf{n}_{ab} + \frac{M_a M_b}{2} \left[ \left( 3(\mathbf{n}_{ab} \cdot \dot{\mathbf{b}})^2 - \dot{\mathbf{b}}^2 \right) \mathbf{n}_{ab} - 2(\mathbf{n}_{ab} \cdot \dot{\mathbf{b}}) \dot{\mathbf{b}} \right] \right]$$

$$+ \frac{1}{r_{ab}^3} \left[ M_a M_b^1 [\delta \mathbf{b} - 3(\delta \mathbf{b} \cdot \mathbf{n}_{ab}) \mathbf{n}_{ab}] - M_b M_a^1 [\delta \mathbf{a} - 3(\delta \mathbf{a} \cdot \mathbf{n}_{ab}) \mathbf{n}_{ab}] \right] \right\},$$

where $M_a \equiv \int_{D_a} \rho \, dV$ and $M_a^1$ [Eq. (A10)] are the 0-order mass of body (*a*) and the PN correction to it; $T_a$, $\varepsilon_a$ and $\tau_a$ are given by Eqs. (A12-13) and (A30); $U^{(a)}$ is that part of the Newtonian potential $U$ which is produced by the bodies $b \neq a$; $G$ is the gravitation constant; $r_{ab} \equiv |\mathbf{a} - \mathbf{b}|$, $\mathbf{n}_{ab} \equiv (\mathbf{a} - \mathbf{b})/r_{ab}$; $\delta \mathbf{a} \equiv \mathbf{a}_1 - \mathbf{a}$, and $\alpha_{ab} \equiv \alpha_a M_b + \alpha_b M_a$ [*cf.* (A8); in Ref. 23, $\alpha_{ab}$ denoted $\alpha_a M_b + \alpha_b M_a - 3\varepsilon_a M_b$]. For applications, it is convenient to change the notation thus:

(4.3) $\quad\quad \mathbf{x}_a \equiv \mathbf{a}, \; \mathbf{x}_{1a} \equiv c^2(\mathbf{a}_{(1)} - \mathbf{a}), \quad\quad \mathbf{x}_{(1)a} \equiv \mathbf{a}_{(1)} = \mathbf{x}_a + \mathbf{x}_{1a}/c^2,$

$\quad\quad\quad\quad \mathbf{u}_a \equiv \dot{\mathbf{a}}, \; \mathbf{u}_{1a} \equiv c^2(\dot{\mathbf{a}}_{(1)} - \dot{\mathbf{a}}) = \dot{\mathbf{x}}_{1a}, \; \mathbf{u}_{(1)a} \equiv \dot{\mathbf{a}}_{(1)} = \mathbf{u}_a + \mathbf{u}_{1a}/c^2,$

[where $\mathbf{a}_{(1)}$ is the *1PN position of the mass center of body* (*a*)] and to rewrite (4.2) in terms of the PN corrections $\mathbf{x}_{1a}$ and $\mathbf{u}_{1a}$ to the positions and velocities:

(4.4) $\quad \dfrac{d\mathbf{u}_{1a}}{dT} = o(\eta^3) + \left[ \left( \dfrac{\tau_a}{M_a} - \dfrac{5}{6} \right) \mathbf{u}_a^2 - \dfrac{5}{3} U^{(a)}(\mathbf{x}_a) - \dfrac{17\varepsilon_a + 6T_a}{3 M_a} \right] \nabla U^{(a)}(\mathbf{x}_a)$

$$- \left[ \left( \frac{\tau_a}{M_a} + 2 \right) \mathbf{u}_a \cdot \nabla U^{(a)}(\mathbf{x}_a) \right] \mathbf{u}_a + \left[ 2 \frac{\varepsilon_a}{M_a} + U^{(a)}(\mathbf{x}_a) \right] \boldsymbol{\omega}_a \wedge \mathbf{u}_a$$

$$+ G \sum_{b \neq a} \left\{ \frac{M_b}{2 r_{ab}} \left[ (\mathbf{n}_{ab} \cdot \dot{\mathbf{u}}_b) \mathbf{n}_{ab} - \dot{\mathbf{u}}_b \right] + \frac{1}{r_{ab}^2} \left[ -\alpha'_{ab} \mathbf{n}_{ab} + \frac{M_b}{2} \left[ \left( 3(\mathbf{n}_{ab} \cdot \mathbf{u}_b)^2 - \mathbf{u}_b^2 \right) \mathbf{n}_{ab} - (\mathbf{n}_{ab} \cdot \mathbf{u}_b) \left( 2 \mathbf{u}_b + \tfrac{8}{3} \mathbf{u}_a \right) \right] \right] \right.$$

$$\left. + \frac{M_b}{r_{ab}^3} \left[ \mathbf{x}_{1b} - \mathbf{x}_{1a} + 3 \left( (\mathbf{x}_{1a} - \mathbf{x}_{1b}) \cdot \mathbf{n}_{ab} \right) \mathbf{n}_{ab} \right] \right\},$$

(4.5) $\quad \alpha'_{ab} \equiv M_b \left( \dfrac{\mathbf{u}_a^2 + \mathbf{u}_b^2}{2} + U^{(a)}(\mathbf{x}_a) + U^{(b)}(\mathbf{x}_b) + \dfrac{11\varepsilon_a + 8T_a}{3 M_a} \right) + M_b^1 + \dfrac{11}{3} \varepsilon_b + \dfrac{8}{3} T_b.$

We note that, in this equation, three *structure-dependent* parameters appear: $T_a$, $\varepsilon_a$, and $\tau_a$.

## 5. THE INITIAL CONDITIONS FOR THE PN CORRECTIONS TO THE MASS CENTERS

The equations of motion are not complete until one has precised the initial conditions. Initial conditions may be imposed on the zero-order approximations for the positions and the velocities of the mass centers, thus

(5.1) $$\mathbf{x}_a(T=0) = \mathbf{x}^0{}_a, \qquad \mathbf{u}_a(T=0) = \mathbf{u}^0{}_a.$$

At first sight, this does not give the initial conditions that should be fulfilled by the first-order positions and velocities, but the definition of the PN mass centers provides relations between the initial zero-order and first-order values. First, [22, Eq. (3.9)$_2$] gives us with [22, Eq. (3.7)$_3$]:

(5.2) $$M_a^1 (\mathbf{a}_1 - \mathbf{a}) = \int_{D_a} (\mathbf{x} - \mathbf{a}) \rho_1(\mathbf{x}) \, dV(\mathbf{x}).$$

(It will be convenient to come back for a while to the notations $\mathbf{a}$, $\dot{\mathbf{a}}$, $\mathbf{a}_1$, $\dot{\mathbf{a}}_1$.) As to the velocities: using the PN mass conservation equations and the definitions, it was proved [22, Eq. (3.21)] that

(5.3) $$M_a^1 \, \dot{\mathbf{a}}_1 = \int_{D_a} (\rho_1 \mathbf{u} + \rho \mathbf{u}_1) \, dV \quad [+ O(1/c^2)],$$

and since the initial conditions for the velocity fields are [22, Eq. (2.34)]

(5.4) $$\mathbf{u}(\mathbf{x},0) = \mathbf{u}_{\text{exact}}(\mathbf{x},0), \qquad \mathbf{u}_1(\mathbf{x},0) = \mathbf{0},$$

we get an equation similar to (5.2):

(5.5) $$M_a^1 (\dot{\mathbf{a}}_1 - \dot{\mathbf{a}}) = \int_{D_a} (\mathbf{u} - \dot{\mathbf{a}}) \rho_1(\mathbf{x}) \, dV(\mathbf{x}) \quad (T=0).$$

It thus remains to calculate the integrals on the r.h.s. of (5.2) and (5.5). Just as $\mathbf{u}_1$ [Eq. (5.4)$_2$], the PN correction $p_1$ to the pressure is zero at the initial time [22, Eq. (2.32)]. Eq. [22, (2.23)] gives

(5.6) $$\rho_1 = (\mathbf{u}^2/2 + U)\rho \quad (T=0).$$

Then, assuming as before rigid internal motions (2.7) and spherical fields $\rho$ inside each body (A16), and using the Taylor expansion of the external potential (A6), we get without difficulty

(5.7) $$M_a^1 (\mathbf{a}_1 - \mathbf{a}) = \gamma_a [\nabla U^{(a)}(\mathbf{a}) - \boldsymbol{\omega}_a \wedge \dot{\mathbf{a}}] + O(\eta^4) \quad (T=0),$$

(5.8) $$M_a^1 (\dot{\mathbf{a}}_1 - \dot{\mathbf{a}}) = \gamma_a [\boldsymbol{\omega}_a \wedge \nabla U^{(a)}(\mathbf{a}) + \boldsymbol{\omega}_a^2 \dot{\mathbf{a}} - (\dot{\mathbf{a}} \cdot \boldsymbol{\omega}_a) \boldsymbol{\omega}_a] + O(\eta^4) \quad (T=0).$$

In the notation (4.3), (5.1), this is

(5.9) $$\mathbf{x}_{1a}(T=0) = \frac{\gamma_a}{M_a} [\nabla U^{(a)}(\mathbf{x}^0{}_a) - \boldsymbol{\omega}_a \wedge \mathbf{u}^0{}_a] + O(\eta^4),$$

$$(5.10) \quad \mathbf{u}_{1a}(T=0) = \frac{\gamma_a}{M_a}[\boldsymbol{\omega}_a \wedge \nabla U^{(a)}(\mathbf{x}^0{}_a) + \omega_a{}^2 \mathbf{u}^0{}_a - (\mathbf{u}^0{}_a \cdot \boldsymbol{\omega}_a)\boldsymbol{\omega}_a] + O(\eta^4).$$

Equation (4.4) for the PN corrections, together with the Newtonian equation and the initial conditions (5.1), (5.9)-(5.10), have been implemented [28-29] in the author's Matlab code [30-31] for the numerical integration and the parameters adjustment in celestial mechanics. (The self-rotation has been neglected for this first implementation, *i.e.*, $\boldsymbol{\omega}_a = \mathbf{0}$.) It has been found that the difference with a standard ephemeris is quite small, thus justifying to truncate the expansions in the separation parameter at the order $\eta^3$ included.

## 6. CONCLUSION

A good separation between extended bodies may be quantified through a small parameter [14], which we denote by $\eta$. Here, we have introduced a natural asymptotic framework for $\eta$, thus allowing to conceptually consider a *family* $S^\eta$ of well-separated systems (each system is weakly gravitating, with the field strength being nearly the same for all systems, and the separation is better and better as $\eta \to 0$). It enables one to define a hierarchy of well-defined approximations for the "tidal" effects, by truncating the expansions with respect to $\eta$. In the absence of a definite asymptotic framework for the good separation in the previous work [23], some numerically significant terms were neglected in the final, explicit equations of motion obtained there. In the present work, all terms up to and including $\eta^3$ have been retained in the PN equations of the scalar theory investigated by the author; the order 3 should be sufficient in the solar system.[3] This method of asymptotic expansions provides an alternative to the method first introduced for test particles [32] and then generalized to extended bodies [5,15], that consists in retaining multipoles up to a certain order. It seems to the author that the present method brings us closer to a numerical control of the error involved in neglecting some terms.

It is also to be noticed that several structure-dependent parameters appear in the final equations of motion (4.4). According to the "asymptotic" method of PN approximation [19-20, 22-24], which has been used in the present work, such appearance seems rather unavoidable, quite independently of the theory considered. This is due to the fact that the general form of the equation for PN corrections [Eq. (4.1) for the investigated theory] involves several integrals of Newtonian fields, which depend on the structure of the body. In another theory, similar equations would be obtained (albeit differing by some coefficients, possibly also by the appearance of new PN gravitational potentials). And, due to the separation between the equations of motion of the order zero and one, which is specific of the asymptotic method, it is hard to see how the structure-dependent integrals could be absorbed in a redefinition of the effective masses (as this turns out to be the case when the standard PN approximation is used [14]). It thus seems likely that structure-dependent parameters should also appear in the final equations of motion in nearly any other theory, when the "asymptotic" method of PN approximation is used.

---

[3] Note that this truncation order means that the main tidal effects *are* included: as is well-known, tidal effects are primarily associated with a space variation in the external gravity acceleration. Thus, such effects are present in the second gradient of the external Newtonian potential, $U^{(a)}{}_{,i,j}$, and the order of the latter is precisely $\eta^3$. But, of course, the main tidal effects affect the Newtonian (zero-order) part of the acceleration, which is standard and has not been explicitly written here.

## APPENDIX: Recalculation of the integrals $M_a^1$, $I^{ai}$, $J^{ai}$ and $K^{ai}$ entering Eq. (4.1)

We shall present in some detail the calculation of the integral $J^{ai} - L^{ai}$, where

(A1) $$L^{ai} \equiv \int_{D_a} \rho \, (\partial^3 W / \partial x^i \, \partial T^2) \, dV,$$

($W$ is given by Eq. (4.14) in Ref. 22), because $J^{ai} - L^{ai}$ contains the terms linear in $\mathbf{a}_1$, that must be retained (Sect. 3). We rewrite Eq. (4.18) of Ref. 22 as

(A2) $$J^{ai} - L^{ai} = \int_{D_a} \sigma_1 U^{(a)}{}_{,i} \, dV + \int_{D_a} \rho \, B^{(a)}{}_{,i} \, dV \quad [+ O(1/c^2)]$$

and, since $\sigma_1$ [22, Eq. (2.24)] and $\rho$ are ord($\eta^0$), we need $U^{(a)}{}_{,i}$ and $B^{(a)}{}_{,i}$ to order $\eta^3$. Because $B = $ N.P.$[\sigma_1]$ [22, Eq. (4.17)], we get

(A3) $$\frac{B^{(a)}(\mathbf{X})}{G} = \sum_{b \neq a} \int_{D_b} \frac{\sigma_1(\mathbf{x})}{|\mathbf{X} - \mathbf{x}|} dV(\mathbf{x}) = \sum_{b \neq a} \int_{D_b} \sigma_1(\mathbf{x}) \left( \varphi(\mathbf{Y}) + \frac{\partial \varphi}{\partial Y^j} h^j + \tfrac{1}{2} \frac{\partial^2 \varphi}{\partial Y^j \partial Y^k} h^j h^k + \ldots \right) dV(\mathbf{x})$$

with $\mathbf{Y} \equiv \mathbf{X} - \mathbf{b}$, $\mathbf{h} \equiv \mathbf{b} - \mathbf{x}$, $\varphi(\mathbf{Y}) \equiv 1/|\mathbf{Y}|$, and $B^{(a)}{}_{,i}$ obtains after differentiating the series term by term with respect to $X^i$. Because any derivative of order $n$ of $1/|\mathbf{Y}|$ has the form $P_n(\mathbf{Y})/|\mathbf{Y}|^{2n+1}$ with $P_n$ a homogeneous polynomial of degree $n$, it follows then from (2.4) that the $n^{\text{th}}$ term in $B^{(a)}{}_{,i}$ is $O(\eta^{n+1})$. Hence we need just the two first terms in the series on the r.h.s. of (A3). We get them from (2.2):

(A4) $$\frac{B^{(a)}(\mathbf{X})}{G} = \sum_{b \neq a} \left( \frac{\alpha_b}{|\mathbf{X} - \mathbf{b}|} + \beta_{bi} \frac{X^i - b^i}{|\mathbf{X} - \mathbf{b}|^3} \right) + O(\eta^3) \qquad (\mathbf{X} \in D_a),$$

(A5) $$\alpha_b \equiv \int_{D_b} \sigma_1(\mathbf{x}) dV(\mathbf{x}), \qquad \beta_{bi} \equiv \int_{D_b} \sigma_1(\mathbf{x})(x^i - b^i) dV(\mathbf{x})$$

(the symbol $\beta_{bi}$ denoted something slightly different in Ref. 23). In the same way, we get successively

(A6) $$U^{(a)}{}_{,i}(\mathbf{X}) = U^{(a)}{}_{,i}(\mathbf{a}) + U^{(a)}{}_{,i,j}(\mathbf{a})(X^j - a^j) + O(\eta^4) \qquad (\mathbf{X} \in D_a),$$

(A7) $$\int_{D_a} \sigma_1 U^{(a)}{}_{,i} \, dV = \alpha_a U^{(a)}{}_{,i}(\mathbf{a}) + \beta_{aj} U^{(a)}{}_{,i,j}(\mathbf{a}) + O(\eta^4).$$

The calculations [23b, Appendix A] give us in the present asymptotic framework for $\eta$:

(A8) $$\alpha_a = M_a[\dot{\mathbf{a}}^2/2 + U^{(a)}(\mathbf{a})] + (8T_a + 11\varepsilon_a)/3 + M_a^1 + O(\eta^3),$$

(A9) $$\beta_{ai} = M_a^1 (a_1^i - a^i) + I^{(a)}{}_{ij} \Omega^{(a)}{}_{jk} \dot{a}^k + \zeta_{ai} + O(\eta^2),$$

(A10) $$M_a^1 = \{M_a[\dot{\mathbf{a}}^2/2 + U^{(a)}(\mathbf{a})] + T_a\}_{T=0} + 2\varepsilon_a + O(\eta^3),$$

in which $\zeta_{ai}$ is defined in Ref. 23, but will cancel as there, and $I^{(a)}{}_{ij}$ is the inertia tensor:

(A11) $$I^{(a)}{}_{ij} \equiv \int_{D_a} \rho (x^i - a^i)(x^j - a^j) \, dV,$$

and where $T_a$ and $\varepsilon_a$, introduced by Fock [14, §74], are defined thus:

(A12) $$T_a \equiv \int_{D_a} \rho \Omega_a \, dV = \Omega^{(a)}{}_{ik} \Omega^{(a)}{}_{jk} I^{(a)}{}_{ij}/2, \quad \Omega_a(\mathbf{x}) \equiv \Omega^{(a)}{}_{ik} \Omega^{(a)}{}_{jk} (x^i - a^i)(x^j - a^j)/2,$$

(A13) $$\varepsilon_a \equiv \int_{D_a} \rho \, u_a \, dV/2 = \int_{\text{space}} (\text{grad } u_a)^2 \, dV/8\pi G.$$

From (A2), (A4) and (A7), it follows that

(A14) $$J^{ai} - L^{ai} = GM_a \frac{\partial}{\partial X^i}\bigg|_{\mathbf{X}=\mathbf{a}} \sum_{b \neq a} \left[ \frac{\alpha_b}{|\mathbf{X}-\mathbf{b}|} + \beta_{bj} \frac{X^j - b^j}{|\mathbf{X}-\mathbf{b}|^3} \right] + \alpha_a \frac{\partial U^{(a)}}{\partial a^i} + \beta_{aj} \frac{\partial^2 U^{(a)}}{\partial a^i \partial a^j} + O(\eta^4).$$

The two terms containing $\beta_{bj}$ and $\beta_{aj}$, respectively, were already found in Ref. 23 [Eq. (2.17) there], but not retained, because both are of order $\eta$ times (any of) the two other terms. In the absence of an asymptotic framework for the parameter $\eta$, we could then only compare *certain couples of terms,* without possibly assigning a definite order in $\eta$ to an individual term, and we decided to retain only the term of the lowest order *in those couples*. Now we can easily evaluate the orders in $\eta$. Thus, we see on (A8)-(A10) that $\alpha_a$ and $\beta_{ai}$ are of order $\eta^0$, hence in (A14) the terms containing $\beta_{bj}$ and $\beta_{aj}$ are ord($\eta^3$), as announced. The assumption of spherical symmetry has *not* been used yet. It gives us

(A15) $$U^{(a)}(\mathbf{x}) = \Sigma_{b \neq a} GM_b/|\mathbf{x} - \mathbf{b}|,$$

(A16) $$I^{(a)}{}_{ij} = \gamma_a \delta_{ij}, \quad \gamma_a \equiv (4\pi/3)\int_0^{r_a} r^4 \rho_a(r) \, dr = \int_{D_a} r^2 \rho \, dV/3 \quad (r \equiv |\mathbf{x}-\mathbf{a}|, \rho(\mathbf{x}) = \rho_a(r), \mathbf{x} \in D_a),$$

(A17) $$\zeta_{ai} = 0.$$

There is no difficulty in adapting the calculation of the integrals $L^{ai}$ and $I^{ai}$ [23b, Appendix A] to ensure that $L^{ai}$ and $\dot{I}^{ai}$ are got at the required order $\eta^3$: using (2.4), (2.8) and (2.10), and considering as before a spherical field $\rho$ in each of the separated bodies, we find that the only change is that the quadratic terms in the rotation velocity (in $I^{ai}$) are not needed any more. Thus:

$$(A18) \quad L^{ai} = -(2/3)\varepsilon_a \, \ddot{a}^i - \frac{GM_a}{2} \sum_{b \neq a} M_b \left( \ddot{b}^k \frac{\partial^2 |\mathbf{a}-\mathbf{b}|}{\partial a^i \partial a^k} - \dot{b}^k \dot{b}^j \frac{\partial^3 |\mathbf{a}-\mathbf{b}|}{\partial a^i \partial a^k \partial a^j} \right) + O(\eta^4),$$

$$(A19) \quad \dot{I}^{ai} = [M_a \dot{\mathbf{a}}^2/2 + 2T_a + 4\varepsilon_a]\ddot{a}^i + M_a \dot{\mathbf{a}} \cdot \ddot{\mathbf{a}}\, \dot{a}^i + M_a \{\ddot{a}^i\, U^{(a)}(\mathbf{a}) + \dot{a}^i \tfrac{d}{dT}[U^{(a)}(\mathbf{a})]\} + o(\eta^3).$$

The recalculation of the integral $K^{ai}$ [23b, Appendix C] is more involved. We have [23]:

$$(A20)\, K^{ai} = \int_{D_a} k_{ij}(p_{,j} + \rho U_{,j})dV + \int_{D_a} p U_{,i} dV + \int_{D_a}(-2U\rho U_{,i})dV - (d/dT)(\int_{D_a} \rho k_{ij} u^j dV)$$
$$+ \int_{D_a} \rho u^j u^k (k_{jk,i} - k_{ik,j}) dV$$

$$\equiv \quad K_1^{ai} \quad + \quad K_2^{ai} \quad + \quad K_3^{ai} \quad - (d/dT) K'^{ai} \quad + \quad K_4^{ai}.$$

One important change is that we need the space tensor $k_{ij} \equiv U h^1{}_{ij}$, hence also $h^1{}_{ij}$ [22, Eq. (2.31)], to the order $\eta^3$ (in Ref. 23 we used the lower-order estimate $h^1{}_{ij} = n_i n_j [1 + O(\eta^2)]$). For spherical fields $\rho$, we get inside body (a):

$$(A21) \quad h^1{}_{ij}(\mathbf{x}) = n_i n_j \left( 1 - \frac{2n_k U^{(a)}{}_{,k}}{u'_a(r)} \right) + \frac{n_i U^{(a)}{}_{,j} + n_j U^{(a)}{}_{,i}}{u'_a(r)} + O(\eta^4), \quad \mathbf{n} \equiv (\mathbf{x}-\mathbf{a})/|\mathbf{x}-\mathbf{a}|$$

(the prime means derivative with respect to $r \equiv |\mathbf{x}-\mathbf{a}|$). Using (2.10) and the following relation of Fock [14, Eq. (73.15)], for which the order of the remainder is easily found:

$$(A22) \quad \rho[u_{a,i} + \Omega^{(a)}{}_{ik} \Omega^{(a)}{}_{jk}(x^j - a^j)] = p_{,i} + O(\eta^3),$$

we simplify a few integrals, e.g. we get after multiplying numerator and denominator by $n_i$:

$$(A23) \quad \int_0^{r_a} r^2 u_a p'/u'_a \, dr = 2\varepsilon_a/4\pi + o(\eta).$$

With the help of this trick, we obtain after somewhat lengthy, but otherwise easy calculations:

$$(A24) \quad K_1^{ai} = [\tfrac{10}{3}\varepsilon_a + \tfrac{5}{3}M_a U^{(a)}(\mathbf{a}) + \tfrac{2}{3}T_a - 2\xi_a]U^{(a)}{}_{,i}(\mathbf{a}) + o(\eta^3),$$

$$(A25) \quad K_2^{ai} = (1/3)(\varepsilon_a - 2T_a) U^{(a)}{}_{,i}(\mathbf{a}) + O(\eta^4),$$

$$\text{(A26)} \qquad K_3{}^{ai} = [2\xi_a - 4\varepsilon_a - 2M_a U^{(a)}(\mathbf{a})]U^{(a)}{}_{,i}(\mathbf{a}) + O(\eta^4),$$

$$\text{(A27)} \qquad \frac{d}{dT}K'{}^{ai} = \frac{d}{dT}\{[\tfrac{2}{3}\varepsilon_a + \tfrac{1}{3} M_a U^{(a)}(\mathbf{a})]\dot{a}^i\} + o(\eta^3),$$

$$\text{(A28)} \qquad K_4{}^{ai} = [M_a U^{(a)}(\mathbf{a}) + 2\varepsilon_a]\Omega^{(a)}{}_{ji}\dot{a}^j + (\tfrac{1}{3}M_a - \tau_a)[\dot{a}^j U^{(a)}{}_{,j}(\mathbf{a})\dot{a}^i - \dot{\mathbf{a}}^2 U^{(a)}{}_{,i}(\mathbf{a})] + o(\eta^3).$$

In these equations, the structure parameters $\xi_a$ and $\tau_a$ are defined thus:

$$\text{(A29)} \qquad \xi_a \equiv -(4\pi/3)\int_0^{r_a}\rho_a u'_a r^3 \, dr,$$

$$\text{(A30)} \qquad \tau_a \equiv [1/(3G)]\int_0^{r_a} u_a\{4r\mu_a'/\mu_a(r) - [r\mu_a'/\mu_a(r)]^2\}dr, \quad \mu_a(r) \equiv 4\pi\int_0^r \rho_a(s)s^2\,ds.$$

[Thus the zero-order mass $M_a$ is just $\mu_a(r_a)$.] These expressions differ from the previously obtained ones [23] only by $O(\eta^3)$ terms, except for $K_1{}^{ai}$. For $K_1{}^{ai}$, there is in (A24) an additional ord($\eta^2$) term: $(8/3)\varepsilon_a U^{(a)}{}_{,i}(\mathbf{a})$, as compared with Ref. 23. The presence of ord($\eta^2$) terms in the new expressions was *a priori* expected since $h^1{}_{ij} - n_i n_j$ is ord($\eta^2$) [see Eq. (A21)]. Thus, the differences with the previous expressions do not reduce to adding higher-order terms. We point out again that this does not come from a calculation error in the previous work, but from the fact that an asymptotic framework for the parameter $\eta$ was missing.

ACKNOWLEDGEMENT

The remarks of the referees allowed to improve the paper.